\documentclass[prl,final,twocolumn,superscriptaddress]{revtex4}
\usepackage{graphicx}

\begin{document}

\title{Scaling behavior in thermoelectric misfit cobalt oxides}

\author{P. Limelette}
\affiliation{Laboratoire LEMA, UMR 6157 CNRS-CEA, Universit\'e F. Rabelais, UFR Sciences, Parc de Grandmont, 37200 Tours, France}
\author{S. H\'ebert}
\affiliation{Laboratoire CRISMAT, UMR 6508 CNRS-ENSICAEN, 6, Boulevard du Mar\'echal Juin, 14050 Caen Cedex, France}
\author{V. Hardy}
\affiliation{Laboratoire CRISMAT, UMR 6508 CNRS-ENSICAEN, 6, Boulevard du Mar\'echal Juin, 14050 Caen Cedex, France}
\author{R. Fr\'esard}
\affiliation{Laboratoire CRISMAT, UMR 6508 CNRS-ENSICAEN, 6, Boulevard du Mar\'echal Juin, 14050 Caen Cedex, France}
\author{Ch. Simon}
\affiliation{Laboratoire CRISMAT, UMR 6508 CNRS-ENSICAEN, 6, Boulevard du Mar\'echal Juin, 14050 Caen Cedex, France}
\author{A. Maignan}
\affiliation{Laboratoire CRISMAT, UMR 6508 CNRS-ENSICAEN, 6, Boulevard du Mar\'echal Juin, 14050 Caen Cedex, France}

\begin{abstract}
\vspace{0.3cm}
We investigate both thermoelectric and thermodynamic properties of the misfit cobalt oxide 
[Bi$_{1.7}$Co$_{0.3}$Ca$_{2}$O$_{4}$]$^{RS}_{0.6}$CoO$_{2}$. 
A large negative magnetothermopower is found to scale with both magnetic field and temperature revealing a significant spin entropy 
contribution to thermoelectric properties giving rise to a constant S$_0\approx$ 60 $\mu$V K$^{-1}$ equal to the 
high temperature asymptotic value of the spin 1/2 entropy. 
Low temperature specific heat measurements allow us to determine an enhanced electronic part with 
$\gamma\approx$ 50 mJ (mol K$^{2}$)$^{-1}$ attesting  of strong correlations. 
Thereby, a critical comparison between [Bi$_{1.7}$Co$_{0.3}$Ca$_{2}$O$_{4}$]$^{RS}_{0.6}$CoO$_{2}$, 
other cobaltites as well as other materials reveals a universal behavior of the thermopower low temperature slope 
as a function of $\gamma$ testifying thus a purely electronic origin. 
This potentially generic scaling behavior suggests here that the high room temperature value of the thermopower in misfit cobalt oxides 
results from the addition of a spin entropy contribution to an enlarged electronic one. 
\end{abstract}

\maketitle
Since the discovery of large thermopower in metallic Na$_{x}$CoO$_2$ and superconductivity in its hydrated form~\cite{Terasaki97,Takada03} the
layered cobalt oxides are at the heart of intense current research. Indeed it was quickly realized that they exhibit striking properties at both low and 
high temperatures, with a large potential for applications. These properties are resulting from an unusual combination of what is conventionally 
interpreted as either metallic or insulating features. Even though their resistivity is metallic like, and exhibits rather low  
value~\cite{Terasaki97,Masset00,Maignan03} as in the so-called {\it bad} metal near the Mott metal-insulator 
transition~\cite{Georges96,Limelette03PRL}, these compounds display a large thermopower (TEP) at room temperature 
(S$_{300 K}\approx  $ 138  $\mu$V K$^{-1}$), which origin is still debated. Increased thermopower usually stems from either strongly enhanced 
effective masses through electronic correlations~\cite{Palsson98,Merino00,behnia04,Motrunich03,LimelettePRB05,Fresard02} or 
from the entropy of electrons localized in degenerate states~\cite{Koshibae00,Koumoto02,Maignan03,Wang03}. 
In most cases one observes that one effects dominates the other. In contrast, we give below evidences that both scenarios contribute to the 
enhanced TEP in misfit cobalt oxides. 

Similarly to Na$_{x}$CoO$_2$, the structure of the incommensurate cobalt oxide 
[Bi$_{1.7}$Co$_{0.3}$Ca$_{2}$O$_{4}$]$^{RS}_{0.6}$CoO$_{2}$ (abbreviated thereafter BiCaCoO) contains single [CoO$_2$] 
layer of CdI$_2$ type stacked with four rocksalt-type layers labeled $RS$ instead of a sodium deficient layer. 
One of the in-plane sublattice parameters being different from one layer to the other~\cite{Maignan03,Leligny99}, the cobaltite 
BiCaCoO is a member of a series of compounds having a misfit structure and high thermopower~\cite{Leligny99,Masset00,Maignan03}.

\begin{figure}[t!]
\centerline{\includegraphics[width=0.95\hsize]{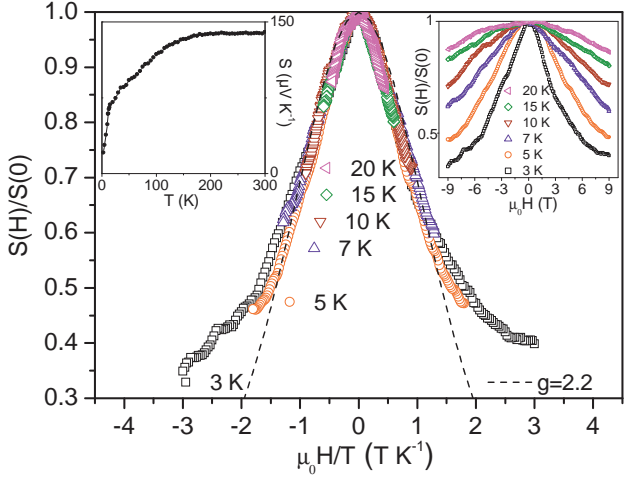}}
\caption{(Color online) Magnetothermopower scaling for a sintered cobalt oxide BiCaCoO. 
The left and the right insets show respectively the temperature and the magnetic field dependences of the 
hole type thermopower (S$>$0). 
The collapse of the data of the right inset as a function of H/T illustrates a spin entropy contribution to the thermopower here normalized to its zero 
magnetic field value for each temperature as S(H)/S(H=0). The dotted line is the free spins 1/2 entropy following Eq.~\ref{eqSspin}.}
\label{figSHT}
\end{figure}

\begin{figure}[htbp]
\centerline{\includegraphics[width=0.95\hsize]{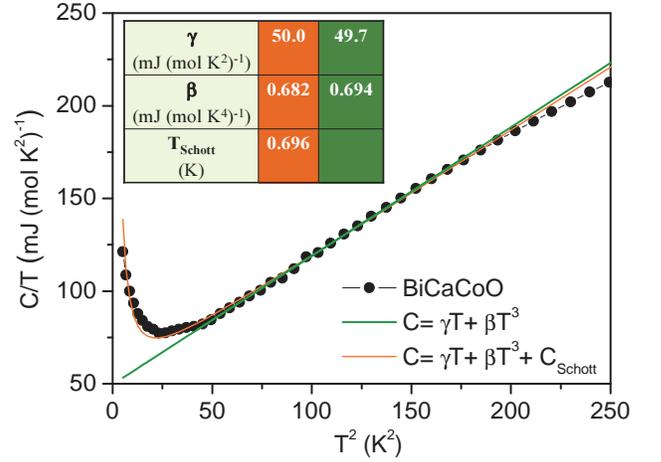}}
\caption{(Color online) Temperature dependence of the specific heat C for a sintered sample as C/T {\it vs.} T$^2$. 
The linear variation of C/T as a function of T$^2$ allows us to determine directly $\gamma \approx$ 50 mJ (mol K$^2$)$^{-1}$.
The inserted table displays the parameters of the two fitting procedures, namely with and without the Schottky term.}
\label{figgamma}
\end{figure}

\begin{figure}[htbp]
\centerline{\includegraphics[width=0.95\hsize]{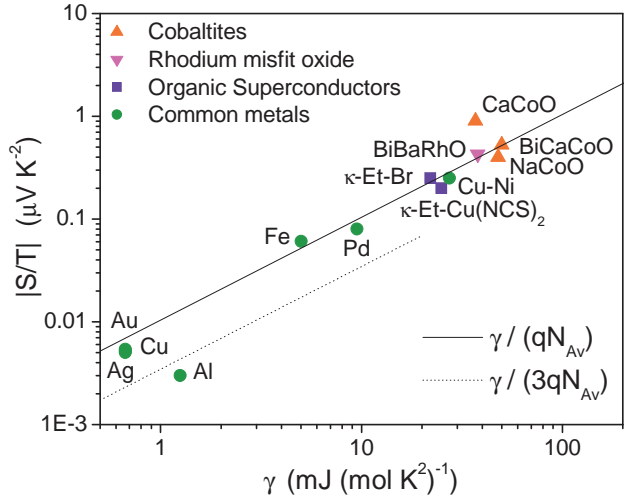}}
\caption{(Color online) Scaling plot of the thermopower slope $|$S/T$|$ as a function of the electronic specific heat coefficient $\gamma$ 
for the cobaltites BiCaCoO, CaCoO~\cite{LimelettePRB05} and NaCoO~\cite{Koumoto02}, the rhodium misfit oxide BiBaRhO~\cite{klein05},
 common metals and organic superconductors~\cite{blatt1960,Ashcroft76,behnia04}. 
The straight and dotted lines are discussed in the text.}
\label{figSgamma}
\end{figure}

We report in this letter on a detailed experimental study of both thermoelectric and thermodynamic properties of the misfit cobalt oxide 
BiCaCoO to determine whether the high room temperature thermopower S could originate from various contributions. 
In order to identify their signatures, we have performed thermopower measurements as a function of magnetic field and temperature. 
As reported in Ref.~\cite{Maignan03} a large negative magnetothermopower is measured at low temperature. 
Here we show that the latter reveals a scaling behavior with both magnetic field and temperature, suggesting a spin entropy contribution 
to the thermopower.\\
Moreover, the determined electronic specific heat coefficient $\gamma \approx$ 50 mJ (mol K$^{2}$)$^{-1}$ leads to a universal ratio S/$\gamma$T independent of interactions that provides a scaling  behavior in a wide range of materials including the two misfit cobalt 
oxides BiCaCoO and [CoCa$_{2}$O$_{3}$]$^{RS}_{0.6}$CoO$_{2}$ (abbreviated CaCoO below). 
Consequently, the thermopower in this misfit cobalt oxide seems to mostly consist of two contributions: 
the first one results from a spin entropy part, and the second one originates likely from strongly renormalized quasiparticles.\\
\indent Thermopower measurement were performed with an experimental setup described in Ref.~\cite{Hejtmanek96} 
on a $2\times 2 \times$ 10 mm$^3$ polycrystalline sintered sample. Fig.~\ref{figSHT} presents the magnetic field (right inset) 
and the temperature (left inset) dependences of the thermopower S. 
As it appears in the left inset of Fig.~\ref{figSHT}, three regimes can be distinguished when the temperature T is varied from 300 K down to 3 K. 
Qualitatively often observed in other misfit cobalt oxides~\cite{Maignan03,Masset00}, the thermopower remains 
first weakly temperature dependent from its room temperature value S$\approx$ 138 $\mu$V K$^{-1}$ 
down to nearly 200 K. Next, S exhibits a quite linear T-dependence down to approximately 20 K, 
followed finally by a strong decrease when the temperature is lowered down to 3 K. 
In the latter temperature range, namely below 20 K, the thermopower shown in the right inset of Fig.~\ref{figSHT} 
reveals a strong magnetic field dependence. A large negative magnetothermopower is observed at 3 K with a 
reduction of about 65 $\%$ between the 0 T and the $\pm$ 9 T values, while a weaker magnetic field effect is seen at 
higher temperatures, nearly vanishing above 20 K. 
\\
\indent The combined influence of both magnetic field and temperature is demonstrated through the complete collapse of the data in 
Fig.~\ref{figSHT} when plotted as a function of H/T. Already observed in the layered cobalt oxide Na$_x$CoO$_2$~\cite{Wang03} 
(abbreviated NaCoO below), this scaling behavior is revealed for the first time in a misfit cobaltite, to our knowledge. 
As mentioned in Ref.~\cite{Wang03},  such a scaling behavior is a typical signature of a spin entropy reduction involving the freeze of spin fluctuations. Thus, we attribute this magnetothermopower to a collection of spins 1/2 
in agreement with the low spin configuration of Co$^{4+}$~\cite{mizokawa05}, with an entropy $\sigma$ related to the thermopower 
as S$_{spin}$=$\sigma$/q, q being the electric charge~\cite{Wang03,Koshibae00}. 
Due to this proportionality, the thermopower acquires the same dependence in H/T as the entropy following Eq.~(\ref{eqSspin}), that holds for free spins 1/2~\cite{Diu89}. 
\begin{equation}
 S(x)/S(0)=\left( \ln{ \left[ 2 \cosh{(x)} \right]} - x\tanh{[x]}  \right)/\ln{(2)}
\label{eqSspin}
\end{equation}
Here x=(g $\mu_B$H)/(2 k$_B$T), g is the Land\'e factor, k$_B$ the Boltzmann constant and $\mu _B$ the Bohr magneton. 
It is worth noting that whereas Eq.~(\ref{eqSspin}) accounts qualitatively for the measured magnetothermopower, 
the quantitative agreement with the experimental results is only efficient with g=2.2 for moderate fields, {\it i.e.} 
when $|\mu_0$H/T$|<$1.5 T K$^{-1}$. 
Nevertheless, we emphasize that the ceramic structure of the studied sample with random orientations of crystallites 
could lower magnetic field effect without altering scaling properties.\\
\indent Thermodynamic measurements have been performed on a classical specific heat option of a Quantum Design PPMS system 
on a sample of 34.7 mg, in order to determine the electronic contribution to the specific heat. Due to the occurrence of a Schottky anomaly, a direct extrapolation of C/T to zero temperature would overestimate the electronic coefficient  $\gamma$, as displayed in Fig.~\ref{figgamma}.\\
\indent For $T \gg T_{\mathrm{Schott}}$ ($T_{\mathrm{Schott}}$ being
the Schottky temperature) the contribution of the Schottky anomaly  
to the specific heat is given 
by $C_{\mathrm{Schott}}\sim$ k$_B$ 
($T_{\mathrm{Schott}}/2T)^2$. Introducing this into the fitting
procedure yields $T_{\mathrm{Schott}} \approx$ 0.696 K, 
and $\gamma \approx $ 50 mJ (mol K$^2$)$^{-1}$. 
Since a direct determination of $\gamma$ from the T$^2$ behavior of C/T observed over a sizeable temperature interval yields the same value 
within our range of accuracy, we confidently conclude that this rather high value suggests an enhanced electronic effective mass m$^*$ which influence upon thermopower is checked below.
\\
\indent As  previously mentioned, one can infer that the linear variation of S within the range 20-125 K in the left inset of Fig.~\ref{figSHT} 
is likely to signal another thermoelectric regime compared to the low temperature regime dominated by a large spin entropy contribution. 
Theoretically, the electronic thermopower is expected to give rise to a linear temperature dependence contribution with a slope S/T 
proportional to the effective mass, namely $\gamma$~\cite{Merino00}. 
Thus, at sufficiently low temperature compared to the Fermi energy, the ratio S/$\gamma$ is predicted in this context to be 
independent of interactions even in a strongly correlated regime as emphasized in Ref.~\cite{Merino00,behnia04}. 
So, in order to check the electronic origin of the linear T-dependence of the thermopower, its slope has been plotted in Fig.~\ref{figSgamma} as 
a function of $\gamma$ in a wide range of materials~\cite{blatt1960,Ashcroft76,behnia04} including the cobaltites as NaCoO~\cite{Koumoto02}, 
CaCoO~\cite{LimelettePRB05}, BiCaCoO and even the rhodium misfit oxide BiBaRhO~\cite{klein05}. 
In analogy with the prototype compound NaCoO, the electronic specific heat coefficients have been considered for each cobaltite per CoO$_{2}$, 
with $\gamma$= 37 mJ (mol K$^2$)$^{-1}$ for CaCoO.
\\
\indent As a matter of fact the linear in $T$ behavior of the thermopower is hardly observed in simple metals, where it is hindered by phononic 
contributions~\cite{Weiss74}. 
In contrast, for both misfit cobaltites, it turns out that its prefactor is sufficiently strongly enhanced by the electronic correlations to yield a large 
temperature domain where the Fermi liquid behavior is observable. 
Thus in this temperature range, the T-dependence of the thermopower is fully governed by the electronic effective mass 
and is renormalized by correlations.\\
\indent In order to give some rather basic theoretical insight into the previous analysis, one may assume following the 
Landau's Fermi liquid theory~\cite{Abriko75} that the main effect of the electronic correlations is to renormalize the effective mass. 
As a consequence, an effective Fermi temperature is defined 
as $T^*\sim T_F m/m^*$~\cite{Georges96}, 
with the bare Fermi temperature T$_F$. In this context, the electronic specific heat coefficient $\gamma^*$ is :
\begin{equation}
\gamma^*= \left( \frac{\pi^2}{2} \right)  \frac{\delta N_{Av} k_B }{ T^*}
\label{eqgamma}
\end{equation}
with the doping $\delta$ and the Avogadro number N$_{Av}$~\cite{Ashcroft76}.
Obviously, one recovers in Eq.~(\ref{eqgamma}) a proportionality between $\gamma^*$ and the effective mass through 1/T$^*$ as expected. 
The electronic thermopower S$^*$ can be inferred in the framework of this basic approach for T$<<$T$^*$ as :
\begin{equation}
S^*= \left( \frac{\pi^2}{6} \right)  \frac{k_B T}{q T^*}
\label{eqS*}
\end{equation}
Since S$^*$ in Eq.~(\ref{eqS*}) is also proportional to the effective mass, it results that the ratio S$^*$/$\gamma^*$T 
in Eq.~(\ref{eqS*gamma}) is thereby independent of interactions as already noted in the context of the dynamical mean field 
theory of strongly correlated systems~\cite{Palsson98,Georges96,Merino00}.
\begin{equation}
 \frac{S^*}{\gamma^* T}= \left( \frac{1}{3 \delta q N_{Av}} \right) 
\label{eqS*gamma}
\end{equation}
Besides, it follows from Eq.~(\ref{eqS*gamma}) that most of the coefficients in Eq.~(\ref{eqgamma}) and Eq.~(\ref{eqS*}) cancel out in 
the ratio S$^*$/$\gamma^*$T that is only doping dependent. 
The latter dependence can thus explain the differences observed in Fig.~\ref{figSgamma} between experimental points 
and the straight line plotted following Eq.~\ref{eqS*gamma} with $\delta$=1/3. 
In particular, the plotted dotted line with $\delta$=1 accounts quite well for aluminium that is characterized by a quasi-spherical 
Fermi surface and one electron per atom. 
So, this analysis suggests that the two misfit cobalt oxides CaCoO and BiCaCoO have a hole doping (S$>$0) lying between nearly 0.13 and 0.33.
\begin{figure}[b!]
\centerline{\includegraphics[width=0.95\hsize]{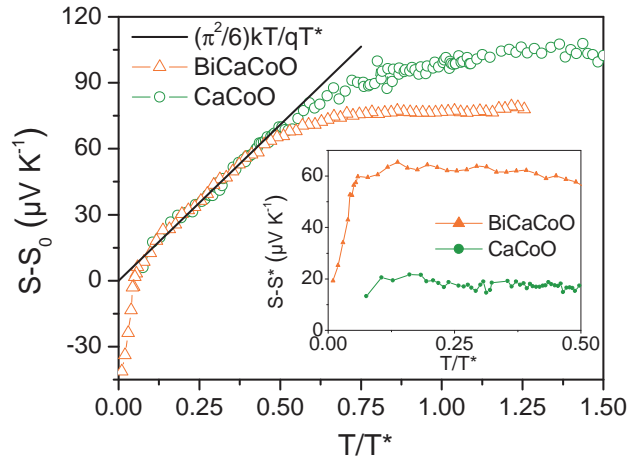}}
\caption{(Color online) Thermopowers (S-S$_0$) as a function of the reduced temperature T/T$^*$, 
with S$_0$ the asymptotic value of the spin entropy contribution and  T$^*$ the effective Fermi temperature. 
A comparison is made between the two cobalt oxides BiCaCoO (S$_0\approx$ 60 $\mu$V K$^{-1}$, T$^*\approx$ 255 K) 
and CaCoO (S$_0\approx$ 20 $\mu$V K$^{-1}$, T$^*\approx$ 140 K) while the straight line shows S$^*$ following Eq.~(\ref{eqS*}).
The inset displays the spin entropy contribution S$_{spin}$=S-S$^*$ for the two cobaltites.}
\label{figSTT}
\end{figure}

Because the relation~(\ref{eqS*}) contains no free parameter besides T$^*$, its direct quantitative comparison with the experimental 
results should fully determine the effective Fermi temperature. 
In order to compare the thermopower kinetic component in both misfit cobaltites, one needs to substract a spin entropy 
contribution in such a way that this component goes linearly with
$T$. 
Above $\sim T^*/10$ this spin entropy contribution looses its
temperature dependence, reaching a value $S_0$ that is substracted
from the data in  Fig.~\ref{figSTT}. As a function of the reduced
temperature T/T$^*$ it shows a collapse of the different curves with
the determined effective Fermi temperatures T$^*_{BiCaCoO}\approx$ 255
K and T$^*_{CaCoO}\approx$ 140 K.  
Moreover, Fig.~\ref{figSTT} clearly displays that the breakdown of the linear dependence of the thermopower seems to occur for the two compounds at the same reduced temperature of the order of T$^*$/2, indicating thus a possible cross-over fully controlled by correlations. 
Finally, we emphasize that while the thermopower is nearly constant above T$^*$, its asymptotic value differs 
from one cobaltite to the other one. 
We stress here that the high temperature regime T$>$T$^*$ involves likely incoherent electronic excitation, namely of higher energy, 
and is thus out of the scope of this low energy analysis. 
It is worth noting that the temperature dependence of the spin entropy contribution S$_{\mathrm{spin}}$ can also be 
extracted from the experimental thermopower subtracting the kinetic contribution defined with the corresponding T$^*$, as S$_{\mathrm{spin}}$=(S-S$^*$). 
Therefore, the spin entropy contribution in the inset of Fig.~\ref{figSTT} yields the constant S$_0\approx$ 20 $\mu$V K$^{-1}$ 
for CaCoO and leads asymptotically to S$_0\approx$ 60 $\mu$V K$^{-1}$ for BiCaCoO. 
One must emphasize that the latter value agrees with the theoretical value S$_0$=k$_B$ln(2)/q$\approx$ 60 $\mu$V K$^{-1}$ 
expected for one free spin 1/2 per Co site in the CoO$_2$ layers.\\
\indent Let us mention that a similar coexistence between delocalized electrons and localized spins has already 
been evidenced in the context of the heavy electron materials within the two fluid Kondo lattice model~\cite{pines2004}. 
While the underlying physics of the latter results from distinct well identified energy scales (Kondo temperature, 
intersite coupling, crystal field effect splitting), here this  coexistence may involve different competing effects. 
First, it is worth  noting that the triangular arrangement of the cobalt sites in this quasi two dimensional structure 
makes the system naturally predisposed to magnetic frustration. 
As emphasized in Ref.~\cite{Motrunich03}, such a system with a strong Coulomb repulsion could be also characterized 
by charge frustration leading to a metallic state with a tendency for short-ranged charge ordering.\\
In addition to this frustration effect, the misfit structure between the [CoO$_2$] and the $RS$ layers may 
introduce disorder by distorting some of the CoO$_6$ octahedrons and lowering the local intersite coupling 
that could also localize electrons.
So for now, in order to investigate both frustration and disorder effects we believe that further X-ray and neutron 
scattering experiments are highly desirable.

To conclude, we have performed both thermopower and thermodynamic measurements in the misfit cobalt oxide BiCaCoO 
focusing on the identification of various regimes. 
Magnetothermopower experiments have demonstrated through a scaling behavior a spin entropy contribution, 
giving rise to a constant S$_0 \approx$ k$_B$ln(2)/q above nearly 20 K. 
While specific heat measurements yield $\gamma \approx $ 50 mJ (mol K$^2$)$^{-1}$, the thermopower T-linear dependence $|S/T|$ 
is found to scale with $\gamma$ in a wide range of materials including the cobaltites BiCaCoO, CaCoO and NaCoO. 
Therefore, we conclude that the thermopower in these misfit cobalt oxides behaves as if it were composed by two components, 
a kinetic one originating from quasiparticles renormalized by electronic correlations and a spin entropy contribution. 
Finally, we would like to address that the ratio S/$\gamma$T seems to provide an efficient probe to check the influence of electronic 
correlations upon thermopower and should be put to experimental test in other misfit cobaltites as well as in other strongly correlated systems.
\begin{acknowledgments}
We are grateful to D. J\'erome,  K. Behnia, G. Kotliar and A. Georges for useful discussions.
\end{acknowledgments}

\end{document}